\pdfoutput=1

\documentclass[11pt]{article}

\usepackage[]{EMNLP2023}

\usepackage{times}
\usepackage{latexsym}

\usepackage[T1]{fontenc}

\usepackage[utf8]{inputenc}

\usepackage{amsthm,amsmath,amssymb}
\usepackage{multirow}
\usepackage{bm}
\usepackage{pgfplots}
\usepackage{tikz}
\usepackage{graphicx}
\usepackage{subcaption}
\usepackage{color}
\usepackage{colortbl}
\usepackage{booktabs}
\usepackage{cleveref}
\usepackage{balance}
\usepackage{makecell}
\usetikzlibrary{patterns}
\usepackage{nicematrix}
\pgfplotsset{compat=1.16}
\usetikzlibrary{pgfplots.groupplots}

\newcommand{\sep}{\left \langle \texttt{SEP} \right \rangle}
\newcommand{\cls}{\left \langle \texttt{CLS} \right \rangle}

\usepackage{pgfkeys}
    \newenvironment{customlegend}[1][]{%
        \begingroup
        \csname pgfplots@init@cleared@structures\endcsname
        \pgfplotsset{#1}%
    }{%
        \csname pgfplots@createlegend\endcsname
        \endgroup
    }%
    \def\addlegendimage{\csname pgfplots@addlegendimage\endcsname}

\definecolor{oceanboatblue}{rgb}{0.0, 0.47, 0.75}

\definecolor{officegreen}{rgb}{0.0, 0.5, 0.0}

\newcommand{\ce}{CO$_2$E}
\usepackage{microtype}

\usepackage{inconsolata}

\pgfplotsset{
    legend image with text/.style={
        legend image code/.code={%
            \node[anchor=center] at (0.3cm,0cm) {#1};
        }
    },
}

\newcommand{\model}{OLEO}

\title{Only Encode Once: Making Content-based News Recommender Greener}

\author{Qijiong Liu \\
  The Hong Kong Polytechnic University \\
  \texttt{liu@qijiong.work} \\ \And
  Jieming Zhu \\
  Huawei Noah's Ark Lab \\
  \texttt{jiemingzhu@ieee.org} \\ \AND
  Quanyu Dai \\
  Huawei Noah's Ark Lab \\
  \texttt{quanyu.dai@connect.polyu.hk} \\ \And
  Xiao-Ming Wu$\thanks{\hspace{0.2cm}Corresponding author.}$ \\
  The Hong Kong Polytechnic University \\
  \texttt{xiao-ming.wu@polyu.edu.hk} \\
}

\begin{document}
\maketitle
\begin{abstract}

Large pretrained language models (PLM) have become de facto news encoders in modern news recommender systems, due to their strong ability in comprehending textual content. These huge Transformer-based architectures, when finetuned on recommendation tasks, can greatly improve news recommendation performance. However, the PLM-based pretrain-finetune framework incurs high computational cost and energy consumption, primarily due to the extensive redundant processing of news encoding during each training epoch. In this paper, we propose the ``Only Encode Once'' framework for news recommendation (OLEO), by decoupling news representation learning from downstream recommendation task learning. The decoupled design makes content-based news recommender as green and efficient as id-based ones, leading to great reduction in computational cost and training resources. Extensive experiments show that our OLEO framework can reduce carbon emissions by up to 13 times compared with the state-of-the-art pretrain-finetune framework and maintain a competitive or even superior performance level. The source code\footnote{\url{https://github.com/Jyonn/GNRS}} is released for reproducibility.

\end{abstract}
\section{Introduction}

\begin{figure}[t]
  \centering
  \includegraphics[width=.9\linewidth]{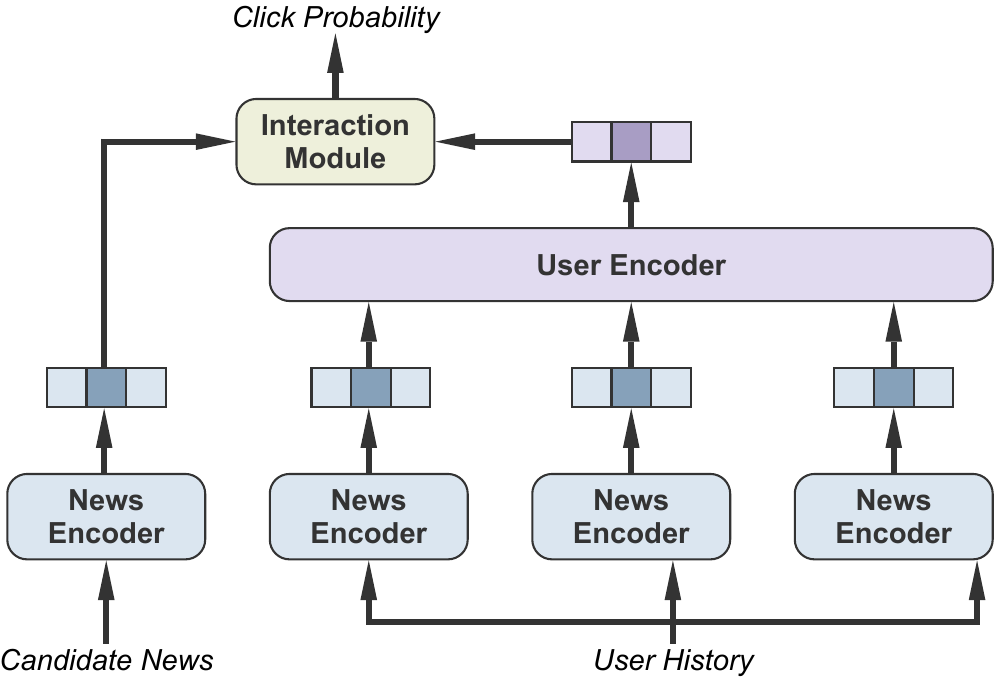}
  \caption{\label{fig:newsrec} 
  A general framework of the news recommender system.
  }
\end{figure}

News recommendation systems have become a crucial tool in the digital age, helping users discover relevant news articles and stay informed amidst the ever-increasing number of sources available. As shown in~\autoref{fig:newsrec}, the general framework of the news recommender system typically consists of three main modules: news encoder, user encoder, and interaction module. This framework accepts a news-user pair as input and yields the click probability. Typically, the news encoder is responsible for capturing the semantic meaning and contextual information of the news articles, while the user encoder processes the news history vectors, capturing the user's preferences, interests, and browsing patterns. Finally, the interaction module facilitates the interaction between the candidate news vector and user vector, ultimately calculating the click probability or relevance score for the given news-user pair.

{
However, this framework is susceptible to redundant processing in the news encoder, resulting in excessive encoding of the same news articles, often repeated thousands of times.} The primary reason behind this redundancy is the frequent occurrence of users in the training process. For instance, in the MIND dataset, an average user appears approximately \textbf{62} times in one training epoch. Moreover, a single news article may appear in multiple user histories. As shown in~\autoref{fig:appearance}, statistical analysis reveals that an average news article appears around \textbf{1818} times within a batch. Redundant processing has not posed significant issues in previous news recommendation systems, primarily due to the simplicity of news encoders, such as using convolutional neural networks~\cite{kim2014convolutional} in NAML~\cite{wu2019neural} or even basic pooling operations in general recommendation models like DCN~\cite{wang2017deep}. {Nevertheless, the redundancy is amplified when more complex network architectures like attention mechanisms or pretrained language models are employed as news encoders}~\cite{zhang2021unbert,wu2021empowering}, as the importance of semantic information in news becomes increasingly recognized. \autoref{fig:param} illustrates the proportion of parameters occupied by different news encoders in various recommendation models. For instance, the PLMNR~\cite{wu2021empowering} model utilizing a 12-layer PLM accounts for over 95\% of the framework's parameters.




\definecolor{lightc}{rgb}{0.86, 0.90, 0.94}

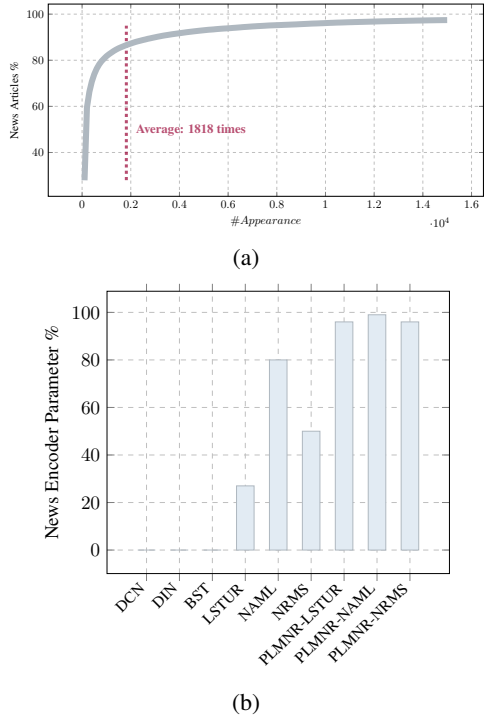
\begin{figure}
    \centering
    \setlength\tabcolsep{0pt}
    \begin{tabular}{c}
    \begin{subfigure}{0.4\textwidth}
        \resizebox{1.0\linewidth}{!}{
            \begin{tikzpicture}
    \begin{axis}[
        ylabel=News Articles \%,
        xlabel=$\#Appearance$,
        line width=0.35mm,
        grid=both,
        grid style=dashed,
        height=8cm,
        width=16cm,
    ]

        \draw[
            purple!80!black,
            dashed,
            line width=1mm,
            opacity=0.7
        ] (1818,28) -- (1818,95) node[right] at (2000, 50) {\large{\textbf{Average: 1818 times}}};
        
        \addplot[
            lightc!80!black,
            line width=2mm,
        ] 
            coordinates{
                (100, 27.94)
                (200, 59.58)
                (300, 66.36)
                (400, 70.75)
                (500, 73.94)
                (600, 76.34)
                (700, 78.17)
                (800, 79.60)
                (900, 80.70)
                (1000, 81.74)
                (1100, 82.57)
                (1200, 83.30)
                (1300, 84.02)
                (1400, 84.64)
                (1500, 85.19)
                (1600, 85.65)
                (1700, 86.08)
                (1800, 86.52)
                (1900, 86.89)
                (2000, 87.25)
                (2100, 87.61)
                (2200, 87.95)
                (2300, 88.24)
                (2400, 88.50)
                (2500, 88.78)
                (2600, 89.02)
                (2700, 89.30)
                (2800, 89.51)
                (2900, 89.75)
                (3000, 89.97)
                (3100, 90.20)
                (3200, 90.40)
                (3300, 90.61)
                (3400, 90.79)
                (3500, 90.95)
                (3600, 91.12)
                (3700, 91.29)
                (3800, 91.43)
                (3900, 91.58)
                (4000, 91.72)
                (4100, 91.87)
                (4200, 92.02)
                (4300, 92.17)
                (4400, 92.30)
                (4500, 92.42)
                (4600, 92.55)
                (4700, 92.67)
                (4800, 92.79)
                (4900, 92.90)
                (5000, 93.01)
                (5100, 93.12)
                (5200, 93.23)
                (5300, 93.30)
                (5400, 93.38)
                (5500, 93.49)
                (5600, 93.58)
                (5700, 93.66)
                (5800, 93.74)
                (5900, 93.80)
                (6000, 93.87)
                (6100, 93.94)
                (6200, 94.03)
                (6300, 94.10)
                (6400, 94.19)
                (6500, 94.27)
                (6600, 94.37)
                (6700, 94.44)
                (6800, 94.51)
                (6900, 94.59)
                (7000, 94.67)
                (7100, 94.73)
                (7200, 94.80)
                (7300, 94.85)
                (7400, 94.91)
                (7500, 94.97)
                (7600, 95.04)
                (7700, 95.10)
                (7800, 95.14)
                (7900, 95.18)
                (8000, 95.23)
                (8100, 95.26)
                (8200, 95.30)
                (8300, 95.35)
                (8400, 95.39)
                (8500, 95.43)
                (8600, 95.48)
                (8700, 95.52)
                (8800, 95.56)
                (8900, 95.62)
                (9000, 95.67)
                (9100, 95.72)
                (9200, 95.77)
                (9300, 95.82)
                (9400, 95.88)
                (9500, 95.92)
                (9600, 95.95)
                (9700, 95.99)
                (9800, 96.04)
                (9900, 96.07)
                (10000, 96.12)
                (10100, 96.17)
                (10200, 96.22)
                (10300, 96.25)
                (10400, 96.30)
                (10500, 96.34)
                (10600, 96.36)
                (10700, 96.39)
                (10800, 96.43)
                (10900, 96.47)
                (11000, 96.50)
                (11100, 96.52)
                (11200, 96.56)
                (11300, 96.58)
                (11400, 96.62)
                (11500, 96.64)
                (11600, 96.67)
                (11700, 96.70)
                (11800, 96.73)
                (11900, 96.75)
                (12000, 96.78)
                (12100, 96.82)
                (12200, 96.85)
                (12300, 96.88)
                (12400, 96.90)
                (12500, 96.93)
                (12600, 96.95)
                (12700, 96.97)
                (12800, 97.00)
                (12900, 97.02)
                (13000, 97.04)
                (13100, 97.07)
                (13200, 97.08)
                (13300, 97.11)
                (13400, 97.14)
                (13500, 97.15)
                (13600, 97.17)
                (13700, 97.20)
                (13800, 97.22)
                (13900, 97.23)
                (14000, 97.26)
                (14100, 97.28)
                (14200, 97.30)
                (14300, 97.32)
                (14400, 97.33)
                (14500, 97.35)
                (14600, 97.37)
                (14700, 97.38)
                (14800, 97.40)
                (14900, 97.41)
                (15000, 97.42)
            };
    \end{axis}
\end{tikzpicture}
        }
        \caption{\label{fig:appearance}}
    \end{subfigure} \\
    \begin{subfigure}{0.35\textwidth}
        \resizebox{1.0\linewidth}{!}{
            \begin{tikzpicture}
\begin{axis}[
    ybar,
    xtick=data,
    xticklabels={\small{DCN}, \small{DIN}, \small{BST}, \small{LSTUR}, \small{NAML}, \small{NRMS}, \small{PLMNR-LSTUR}, \small{PLMNR-NAML}, \small{PLMNR-NRMS}},
    ylabel=News Encoder Parameter \%,
    enlarge x limits=0.15,
    xticklabel style={rotate=45, anchor=east},
    grid=both,
    grid style=dashed,
]
\addplot[lightc!80!black, fill=lightc!80!white] coordinates {(1,0) (2,0) (3,0) (4,27) (5,80) (6,50) (7,96) (8,99) (9,96)};
\end{axis}
\end{tikzpicture}
        }
        \caption{\label{fig:param}}
    \end{subfigure}
    \end{tabular}
    
    \caption{(a) The relation between news appearance times and proportion of news articles. Each point (x, y) represents y \% of news articles appear less than x times within a training epoch. The line ends at the point (550700, 100). (b) Proportion of parameters occupied by different news encoders in various recommendation models.}
\end{figure}

\begin{figure}[t]
  \centering
  \includegraphics[width=.65\linewidth]{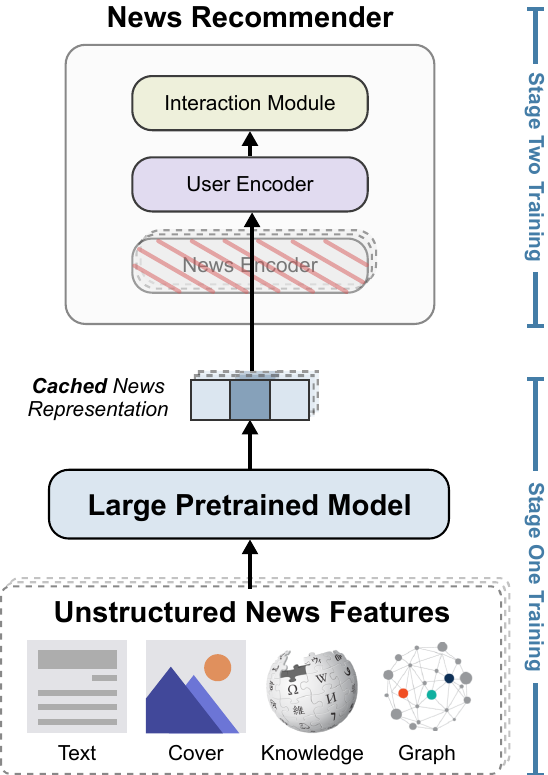}
  \caption{\label{fig:intro} 
  Proposed Two-stage \model{} Framework.
  }
\end{figure}

To alleviate the redundancy processing in news encoder and enhance the training and inference efficiency of the recommendation system, we propose a novel ``\textbf{O}n\textbf{l}y \textbf{E}ncode \textbf{O}nce'' framework, namely \textbf{\model{}}. \model{} decouples the original framework into two distinct stages: \textbf{news representation learning (NRL)} and downstream recommendation task learning.
As shown in \autoref{fig:intro}, the first stage of our proposed framework, known as the News Representation Learning (NRL) module, is dedicated to comprehending the content of news articles and extracting their representations through self-supervised learning. Importantly, this stage operates independently of downstream recommendation tasks such as matching or ranking.

{This decoupling results in several advantages in terms of resource utilization.} Firstly, the NRL module only needs to be trained \emph{\textbf{once}} and can be utilized across $N$ different recommendation models. In contrast, traditional coupled frameworks would require training $N$ times. Secondly, in the NRL stage, each news article appears only \emph{\textbf{once}} during a single training epoch, eliminating the previous issue of an article appearing excessively (1818 times).
Subsequently, during the second stage of recommendation task learning, only the user encoder and interaction module are trained, resembling id-based sequential recommenders. However, these components can also leverage the content features extracted by the NRL module.
Hence, our \model{} framework ensures the model efficiency akin to id-based recommenders, while simultaneously leveraging the abundant semantic information found in content-based recommenders, aligning with the goal of Green AI~\cite{schwartz2020green}.

To validate the efficacy of our \model{} framework, we instantiate the NRL module with a simple yet efficient multi-field Transformer (MFT) architecture. Additionally, we introduce two in-domain pretraining tasks specifically designed for training the MFT on news datasets and extracting meaningful news representations. 
To summarize, this paper makes the following key contributions:

1) We propose a novel news recommendation framework \model{}, which optimizes resource utilization and incorporates task-agnostic and model-agnostic news representation learning. Our \model{} framework aims to promote the development of energy-efficient news recommender systems, contributing to the advancement of Green AI and serving as a guide for future research in this domain.

2) We propose a specific instantiation of news representation learning using a self-supervised multi-field Transformer (MFT). This design enables the capturing of rich semantics in news articles and the learning of informative news representations. Through extensive experiments, we demonstrate the effectiveness and energy efficiency of MFT, which achieves a favorable tradeoff between recommendation performance and energy utilization compared to state-of-the-art methods.

3) We conduct a comprehensive benchmarking and evaluation of popular news recommendation models using both existing training frameworks and our proposed \model{} framework. These benchmark results serve as robust baselines for future research. Additionally, we will make the news representation vectors pretrained on the MIND dataset~\cite{wu2020mind} and the evaluation code for downstream tasks publicly available as benchmarks for the community.

\section{Proposed Framework: \model{}}

Our proposed \model{} framework is illustrated in \autoref{fig:intro}.  In this section, we elaborate on the components of \model{} including news representation learning, downstream recommendation tasks, and evaluation protocols.

\subsection{News Representation Learning (NRL) }

Understanding news content is a critical yet challenging task in news recommendation. Existing approaches address this by either employing advanced network structures, such as CNNs~\cite{wu2019npa} or self-attentions~\cite{wu2019nrms}, to learn news representations in an end-to-end training manner, or by utilizing pretrained language models like BERT~\cite{devlin2018bert} within a pretrain-finetune framework.
However, as the size of news encoder grows, these coupled models tend to suffer from slow training and inference times due to the redundant processing in news encoding. This limitation significantly hinders their practical adoption.
Instead, our work introduces the news representation learning task, which focuses on offline self-supervised learning of news representations.By training the news representations in advance, we can freeze and cache them for downstream recommendation tasks, leading to more practical usage of NRL.

Given the unstructured news features (e.g., text, cover image, entities), news representation learning poses a number of new challenges and opportunities compared to traditional text representation tasks that have been widely studied in the NLP community. In pursuit of effective news representation learning, we identify several potential directions for future research:

\textbf{Text-based NRL.} 
Applying state-of-the-art NLP techniques to understand news texts efficiently for recommendation tasks remains a significant challenge. Previous work~\cite{zhang2021unbert,wu2021empowering} has utilized pretrained language models but faces inefficiency issues, highlighting the need for efficient universal representation learning from news texts.


\textbf{Multi-field NRL.} 
News articles consist of various data fields, including title, abstract, body, category, keywords, and publisher. Leveraging multi-field information has shown promise in improving end-to-end news recommendation models~\cite{wu2019neural}. However, effectively utilizing such semi-structured information in a self-supervised manner for better news representation remains an open research question.

\textbf{Multi-modal NRL.} With the prevalence of multi-modal online news containing textual contents and cover images, modeling visual information becomes crucial in news representation learning to capture users' visual interests. Exploring the integration of pretrained multi-modal models, like CLIP~\cite{CLIP}, into news representation learning for downstream recommendation is a promising research direction.

\textbf{Knowledge-enhanced NRL.} 
News articles often contain entities that can be connected in a knowledge graph. There exist some pioneer studies on how to train a pretrained language model with knowledge enhancement~\cite{EnhancingKnowledge}. However, incorporating such knowledge entities in news representation learning, along with textual and visual information, poses a demanding research topic, particularly for recommendation tasks.

\textbf{Graph-enhanced NRL.} 
News representation learning can benefit from fusing information through a graph view. For example, news articles can be connected through entities in a knowledge graph or co-occurrence graph derived from historical data. Exploring the effective utilization of graph structures in news representation learning is an important research question to address.

\subsection{Downstream Recommendation Tasks}

Downstream recommendation tasks can vary according to different objectives and application scenarios. In this work, we choose 6 widely-used news recommendation models with 5 variants (30 in total) for evaluation, including 4 matching models (i.e., NAML~\cite{wu2019neural}, LSTUR~\cite{an2019neural}, NRMS~\cite{wu2019nrms}, and BST~\cite{bst}), and 2 ranking models (i.e., DCN~\cite{wang2017deep} and DIN~\cite{zhou2018deep}. Variants are under end-to-end framework, pretrain-finetune framework, and our proposed \model{} framework. More details please refer to~\autoref{subsec:baseline}.

\begin{figure*}[t]
\centering
\includegraphics[width=.8\linewidth]{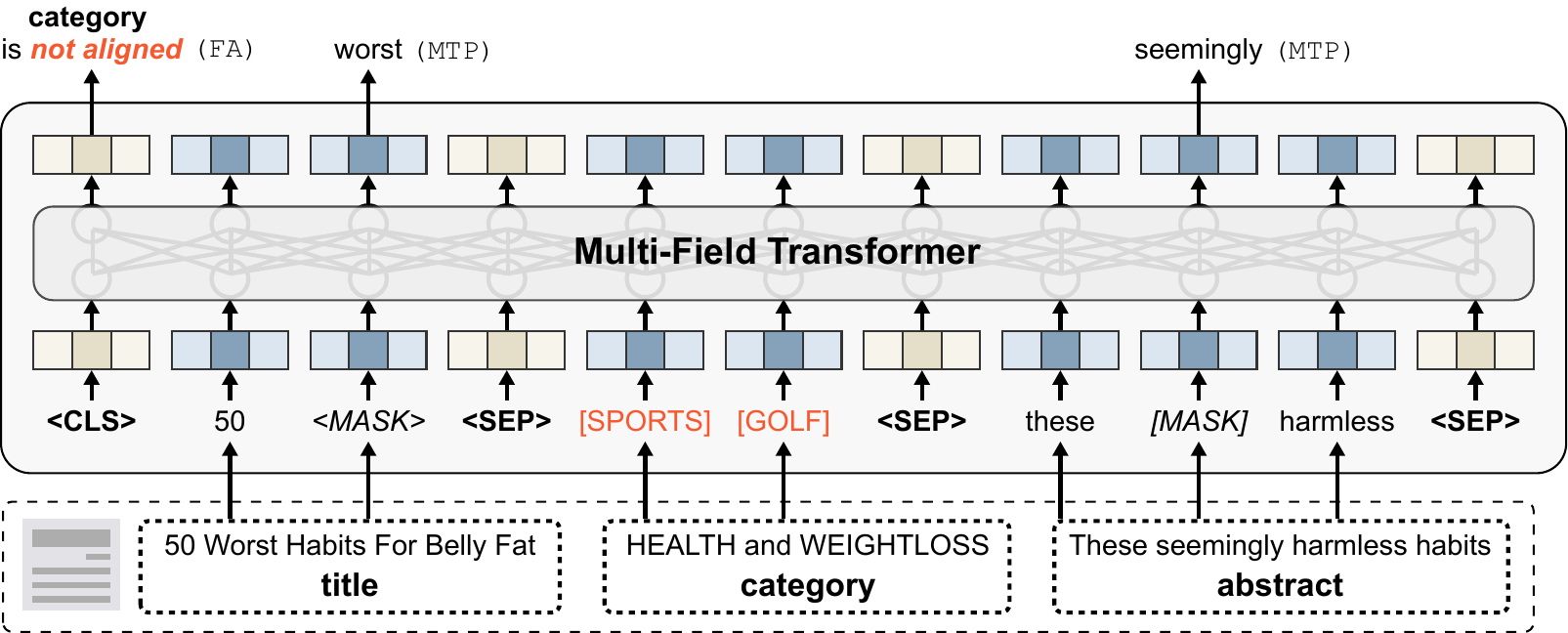}
\caption{\label{fig:overview} Overview of our proposed multi-field Transformer and two self-supervised tasks, i.e., masked token prediction (MTP) and field alignment (FA).}
\end{figure*}

\subsection{Representation Evaluation Protocol}\label{subsec:protocal}
The problem definition of news representation learning is inspired by the widely studied research topic on self-supervised visual representation learning~\cite{VisualSelfSupervision}. In the CV domain, the learned image representations are frozen and evaluated through an evaluation protocol of linear classification or MLP-like classification~\cite{VisualSelfSupervision}. 
Likewise, we apply the widely used NRMS~\cite{wu2019nrms}) and DCN~\cite{wang2017deep} for downstream matching and ranking tasks, respectively. Formally, we define a standard two-stage evaluation protocol as follows.

1) A NRL task aims to learn representations from news content through self-supervision tasks:
\begin{equation}
    h = \Phi(x)~,
\end{equation}
where $\Phi$ denotes the news encoder, which takes the raw content $x$ as input and outputs the news representation vector $h$. $\Phi$ is trained through self-supervision tasks. Instead of applying the pretraining-finetuning paradigm that largely slows down the inference speed of downstream models, $\Phi$ is frozen after training and used to obtain representations of all news in the corpus for downstream tasks.

2) A downstream recommendation model, which takes the learned representations as input, is trained to predict the matching score or click probability of a given user-news pair.
\begin{equation}
    \hat{y} = \Psi(h)~,
\end{equation}
where $h$ is the input representation and $\Psi$ denotes one of the downstream model architecture with fixed network sizes to fairly compare against different representations. $\hat{y}$ denotes the matching score or predicted click probability. 
We evaluate the accuracy of model $\Psi$ in terms of the common evaluation metrics, including AUC, MRR, nDCG@5~\cite{wu2020mind}.
We also introduce a Green AI metrics: Carbon Emission~\cite{lacoste2019quantifying} (\ce) and AUC per Carbon Emission (ApC), to evaluate the sustainability and energy utilization of model $\Psi$. $CO_2E$ is computed by:
\begin{equation}
    CO_2E = p \times t \times c,
\end{equation}
where $p$ is the power consumption differ by hardware types, $t$ is the hardware running time, and $c$ is carbon efficiency differ by power provider. ApC measures the unit conversion rate of carbon emissions on AUC, defined as:
\begin{equation}
    ApC = \frac{AUC - 50}{CO_2E} \times 100.
\end{equation}

\section{Instantiation}

\subsection{Overview}

In this section, we present our initial approach to news representation learning, focusing on providing effective representations for downstream recommendation models. Inspired by PREC~\cite{prec}, our method employs a multi-field Transformer to capture knowledge from multi-fields news content.

For simplicity, we assume that news articles consist of three fields: title, abstract, and category. However, our approach can easily be extended to incorporate additional fields. We denote the title, category, and abstract fields as $\textbf{t}$, $\textbf{c}$, and $\textbf{a}$ respectively. The sequence of field $i$ is represented as $\bm s_i = [s_{i,1}, ..., s_{i,|\bm s_i|}]$, where $i \in \left \{ \textbf{t}, \textbf{c}, \textbf{a} \right \}$ and $|\bm s_i|$ represents the length of the field.


\subsection{Multi-field Transformer}

As shown in \autoref{fig:overview}, the Multi-field Transformer (MFT) takes in a sequence of multi-field news content, where each field is marked with a $\sep$ identifier. The sequence is defined as:
\begin{equation}
    \bm s = \left[\cls, s_{t,1},\cdot,\sep,s_{c,1},\cdot,\sep,s_{a,1},...\sep\right].
\end{equation}
Following BERT, the input embedding consists of token embedding, position embedding, and segment embedding, which are integrated as follows:
\begin{gather}
    \bm E^0 = \bm E_\text{token} + \bm E_\text{pos} + \bm E_\text{field} \in \mathbb{R}^{m \times d},
\end{gather}
where $m = |\bm s_t| + |\bm s_c| + |\bm s_a| + 4$ is the sequence length, and $d$ is the latent embedding dimension.

MFT consists of $H$ Transformer layers.
In each layer, the input sequence $\bm E^{i-1}$ is transformed by:
\begin{equation}
    \bm E^{i} = {Trm} \left(\bm E^{i-1}\right), i=1,2,...,H,
\end{equation}
Through multiple Transformer layers, the output sequence $\bm E^H$ contains refined and contextualized representations for each token. An average pooling operation is then applied to aggregate token representations into a unified news representation.

\subsection{Multi-level Self-supervised Tasks}

To learn news representations, we design two in-domain self-supervised tasks at token-level and field-level, as illustrated in \autoref{fig:overview}.

\textbf{Masked Token Prediction (MTP)} is a token-level task inspired by the masked language modeling task in BERT. It aims to predict masked tokens based on contextual information. The loss function is defined as:
\begin{gather}
    L_\texttt{MTP} = -\frac{1}{n} \sum^n_{i=1} \sum_{\bm z \in \bm s} \log \left(\texttt{C}_{mtp}\left(\bm e\right)_{v}\right), \\
    \texttt{C}_{mtp}: \mathbb{R}^d \rightarrow \mathbb{R}^N,
\end{gather}
where $\bm z$ is the $\left(v, \bm e\right)$ pair (i.e., a masked item and its corresponding latent embedding) in sample $\bm s$, $n$ is the batch size, and $N$ is the vocabulary size.

\textbf{Field Alignment (FA)} is a field-level task that aims to capture inter-field relations. Positive samples consist of the original news articles, while negative samples are created by replacing a certain field of the news. The task is to minimize the following loss:
\begin{gather}
    L_\texttt{FA} = -\frac{1}{n} \sum^n_{i=1} \sum_{\bm z \in \bm s} \log \left(\texttt{C}_{fa}\left(\bm e\right)_{l}\right), \\
    \texttt{C}_{fa}: \mathbb{R}^d \rightarrow \mathbb{R}^2,
\end{gather}
where $\bm z$ represents the $\left(l, \bm e\right)$ pair, and $l \in \left\{0, 1\right\}$ is the alignment label.

\newcommand{\mosmall}[1]{\texttt{\textcolor[HTML]{3E7A3E}{#1}}}  
\newcommand{\momedian}[1]{\texttt{\textcolor[HTML]{6C75D6}{#1}}}  
\newcommand{\molarge}[1]{\texttt{\textcolor[HTML]{DB3636}{#1}}}  

\begin{table*}[t]

\centering
\renewcommand\arraystretch{1.2}
\setlength\tabcolsep{4pt}

\resizebox{.9\linewidth}{!}{

\begin{tabular}{llllllll}
\toprule
 & \multicolumn{1}{c}{\textbf{ID-based}} & \multicolumn{3}{c}{\textbf{End-to-end Text-based}} & \multicolumn{1}{c}{\textbf{Pretrain-Finetune}} & \multicolumn{2}{c}{\textbf{\model{}}} \\
\cmidrule(lr){2-2} \cmidrule(lr){3-5} \cmidrule(lr){6-6} \cmidrule(lr){7-8} 
 & \textbf{Null} & \textbf{Pooling} & \textbf{CNN}~\citeyearpar{krizhevsky2012imagenet} & \textbf{Attention}~\citeyearpar{vaswani2017attention} & \textbf{BERT}~\citeyearpar{devlin2018bert} & \textbf{BERT}~\citeyearpar{devlin2018bert} & \textbf{MFT} \textit{(ours)} \\
\midrule
\textbf{Null} & \makecell[l]{DIN~\citeyearpar{zhou2018deep}\\\mosmall{6.91M}} & \makecell[l]{Text-DIN\\\momedian{28.59M}} & - & - & \makecell[l]{PLMNR-DIN\\\molarge{52.02M}} & \makecell[l]{BERT-DIN\\\mosmall{6.91M}} & \makecell[l]{MFT-DIN\\\mosmall{6.91M}} \\
\textbf{Pooling} & \makecell[l]{DCN~\citeyearpar{wang2017deep}\\\mosmall{5.32M}} & \makecell[l]{Text-DCN\\\momedian{27.01M}} & - & - & \makecell[l]{PLMNR-DCN\\\molarge{50.44M}} & \makecell[l]{BERT-DCN\\\mosmall{5.32M}} & \makecell[l]{MFT-DCN\\\mosmall{5.32M}}\\
\makecell[l]{\textbf{Additive}\\\textbf{Attention}~\citeyearpar{bahdanau2014neural}} & \makecell[l]{ID-NAML\\\mosmall{2.36M}} & - & \makecell[l]{NAML~\citeyearpar{wu2019neural}\\\momedian{26.41M}} & - & \makecell[l]{PLMNR-NAML~\citeyearpar{wu2021empowering}\\\molarge{47.48M}} & \makecell[l]{BERT-NAML\\\mosmall{2.36M}} & \makecell[l]{MFT-NAML\\\mosmall{2.36M}}\\
\textbf{GRU}~\citeyearpar{chung2014empirical} & \makecell[l]{ID-LSTUR\\\mosmall{12.99M}} & - & \makecell[l]{LSTUR~\citeyearpar{an2019neural}\\\momedian{32.31M}} & - & \makecell[l]{PLMNR-LSTUR~\citeyearpar{wu2021empowering}\\\molarge{51.02M}} & \makecell[l]{BERT-LSTUR\\\mosmall{12.99M}} & \makecell[l]{MFT-LSTUR\\\mosmall{12.99M}} \\
\textbf{Attention}~\citeyearpar{vaswani2017attention} & \makecell[l]{ID-NRMS\\\mosmall{5.32M}} & - & - & \makecell[l]{NRMS~\citeyearpar{wu2019nrms}\\\momedian{30.55M}} & \makecell[l]{PLMNR-NRMS~\citeyearpar{wu2021empowering}\\\molarge{50.43M}} & \makecell[l]{BERT-NRMS\\\mosmall{5.32M}} & \makecell[l]{MFT-NRMS\\\mosmall{5.32M}} \\
\textbf{Transformer}~\citeyearpar{vaswani2017attention} & \makecell[l]{BST~\citeyearpar{bst}\\\mosmall{25.20M}} & \makecell[l]{Text-BST\\\momedian{46.89M}} & - & - & \makecell[l]{PLMNR-BST\\\molarge{70.32M}} & \makecell[l]{BERT-BST\\\mosmall{25.20M}} & \makecell[l]{MFT-BST\\\mosmall{25.20M}} \\
\bottomrule
\end{tabular}

}

\caption{Methods and variants that will be benchmarked and compared. The first row is a set of different news encoders, and the first column is a set of different user encoders.\label{tab:benchmark}}
\end{table*}
\begin{table}[t]
\centering
\renewcommand\arraystretch{1.2}

\resizebox{.85\linewidth}{!}{
\begin{tabular}{ccc}
\toprule
 & \textbf{MIND-small} & \textbf{MIND-large} \\ \hline
$\#$ \textbf{News} & 65,238 & 104,151 \\
$\#$ \textbf{Users} & 94,057 & 750,434 \\
$\#$ \textbf{Interactions} & 347,727 & 3,958,501 \\
$\#$ \textbf{Samples} & 8,381,093 & 95,447,571 \\
\textbf{Density} & 0.1366\% & 0.1221\% \\
\bottomrule
\end{tabular}
}
\caption{\label{tab:dataset} Dataset statistics.}
\end{table}



The overall optimization objective is defined by:
\begin{equation}
    L = L_{\texttt{MTP}} + L_{\texttt{FA}}.
\end{equation}

\section{Experiments}


\subsection{Experimental Setup}

Our experiments are conducted on a large-scale English news recommendation dataset called MIND (Microsoft News Dataset)~\cite{wu2020mind}. We evaluate all methods on both the MIND-large and MIND-small versions of the dataset. To ensure a fair evaluation, we divide the validation dataset equally into separate validation and test datasets. It's worth noting that the previous work did not provide clear information about the division method of MIND-small, and the test dataset of MIND-large was not disclosed and required online testing. For the convenience of future benchmarking, we have re-partitioned the validation and test sets while keeping the training set unchanged. The detailed statistics of the dataset are presented in \autoref{tab:dataset}. During the pretraining of MFT, we randomly split the news set into a 4:1 ratio, with 80\% for training and 20\% for validation purposes.

\textbf{Evaluation Protocols.} 
We evaluate our method in terms of ranking performance (with metrics as AUC, MRR, and nDCG) and sustainability (with metrics as CO$_2$E and ApC). 
Please refer to~\autoref{subsec:protocal} for details.

\textbf{Implementation Details.} We utilize the BertTokenizer from the transformers library~\cite{wolf-etal-2020-transformers} to tokenize the news content. The Adam optimizer is employed for optimization. We consider various learning rates from the set $\{1e-5, 2e-5, 5e-5, 1e-4, 2e-4, 5e-4\}$ and batch sizes from the set $\{64, 128, 256, 500, 1000, 5000\}$. The embedding dimensions for all baseline methods' variants are fixed at 64. In the case where the model is initialized with news representations, we keep the pretrained embeddings fixed and use learnable transformation matrices for dimension projection. The number of MFT transformer layers is set to 3. We have made the code publicly available with comprehensive details. Experimental results are averaged over five runs, and all methods were trained using Nvidia GeForce RTX 3090 with 24GB memory, which has a power consumption of 350W. The carbon efficiency in our experimental region (anonymous for review) is 722g CO2-eq/kWh. The total computational budget utilized was less than 2000 GPU hours.

\newcommand{\coesmall}[1]{\textbf{\textcolor[HTML]{3E7A3E}{#1}}}
\newcommand{\coemedian}[1]{\textbf{\textcolor[HTML]{6C75D6}{#1}}}
\newcommand{\coelarge}[1]{\textbf{\textcolor[HTML]{DB3636}{#1}}}

\begin{table*}[t]

\centering
\renewcommand\arraystretch{1.2}

\resizebox{\linewidth}{!}{
\begin{tabular}{llcccccccccccc}
\toprule
\multicolumn{2}{c}{\textbf{Dataset}}  & \multicolumn{6}{c}{\textbf{MIND-small}} & \multicolumn{6}{c}{\textbf{MIND-large}} \\
\cmidrule(lr){3-8} \cmidrule(lr){10-14}
&  & \multicolumn{4}{c}{\textbf{Matching}} & \multicolumn{2}{c}{\textbf{Ranking}} & \multicolumn{4}{c}{\textbf{Matching}} & \multicolumn{2}{c}{\textbf{Ranking}} \\
\cmidrule(lr){3-6} \cmidrule(lr){7-8} \cmidrule(lr){9-12} \cmidrule(lr){13-14}
\multicolumn{2}{c}{\textbf{Method}} & \textbf{NAML} & \textbf{LSTUR} & \textbf{NRMS} & \textbf{BST} & \textbf{DCN} & \textbf{DIN} & \textbf{NAML} & \textbf{LSTUR} & \textbf{NRMS} & \textbf{BST} & \textbf{DCN} & \textbf{DIN} \\
\midrule
\multirow{5}{*}{\coesmall{ID-based}}
 & \textbf{AUC} & 50.13 & 51.04 & 54.84 & 50.09 
       & 53.92 & 55.95 
       & 52.98 & 54.98 & 57.59 & 52.10 
       & 57.41 & 57.36 \\
 & \textbf{MRR} & 23.01 & 22.90 & 26.53 & 22.13 
       & 25.18 & 25.88 
       & 24.52 & 25.99 & 27.41 & 24.81 
       & 26.76 & 26.70 \\
 & \textbf{N@5} & 22.35 & 22.31 & 26.34 & 21.59 
       & 24.43 & 25.95 
       & 24.12 & 25.64 & 27.05 & 24.63 
       & 26.90 & 26.84 \\
 & \textbf{\ce $\downarrow$} & \textbf{19}    & \textbf{20}    & \textbf{28}    & \textbf{38}    
       & \textbf{50}    & \textbf{58}    
       & \textbf{294}   & \textbf{353}   & \textbf{471}   & \textbf{555}   
       & \textbf{791}   & \textbf{926}   \\
 & \textbf{ApC} & 0.68  & 5.20  & 17.29 & 0.24  
       & 7.84  & 10.26 
       & 1.01  & 1.41  & 1.61  & 0.38  
       & 0.94  & 0.79  \\
\midrule
\multirow{5}{*}{\coemedian{Text-based}} 
 & \textbf{AUC} & 60.14 & 61.27 & 62.21 & 60.51 
       & 62.63 & 62.90 
       & 63.03 & 63.89 & 64.12 & 63.28 
       & 63.88 & 64.02 \\
 & \textbf{MRR} & 28.93 & 29.64 & 30.19 & 28.59 
       & 29.73 & 30.06 
       & 30.40 & 31.24 & 31.77 & 30.73 
       & 31.65 & 31.98 \\
 & \textbf{N@5} & 29.33 & 30.28 & 31.10 & 29.09 
       & 30.52 & 30.65 
       & 31.82 & 32.15 & 32.64 & 31.95 
       & 32.40 & 33.00 \\
 & \textbf{\ce $\downarrow$} & 42    & 58    & 62    & 53    
       & 75    & 79    
       & 648   & 892   & 1010  & 1212  
       & 1684  & 1746  \\
 & \textbf{ApC} & 24.14 & 19.43 & 19.69 & 19.83 
       & 16.84 & 16.33 
       & 2.01  & 1.56  & 1.40  & 1.09  
       & 0.82  & 0.80  \\
\midrule
\multirow{5}{*}{\coelarge{PLMNR}}
 & \textbf{AUC} & \underline{62.06} & \textbf{63.64} & \underline{62.53} & \textbf{64.40} 
       & \underline{63.32} & \textbf{63.26} 
       & \textbf{65.19} & \textbf{65.73} & \textbf{65.57} & \textbf{66.03} 
       & \underline{65.42} & \textbf{65.31} \\
 & \textbf{MRR} & \textbf{31.66} & \textbf{31.74} & \underline{30.74} & \textbf{32.21} 
       & \underline{32.00} & \textbf{31.83} 
       & \textbf{32.74} & \textbf{33.18} & \textbf{32.94} & \textbf{33.40} 
       & \underline{32.85} & \underline{32.68} \\
 & \textbf{N@5} & \textbf{32.25} & \textbf{32.72} & \underline{31.31} & \textbf{33.34} 
       & \underline{32.58} & \textbf{32.40} 
       & \textbf{33.77} & \textbf{34.26} & \textbf{34.13} & \textbf{34.70} 
       & \underline{33.99} & \textbf{33.70} \\
 & \textbf{\ce $\downarrow$} & 178   & 202   & 252   & 505   
       & 277   & 606   
       & 2527  & 3032  & 4043  & 8086  
       & 6570  & 10108 \\
 & \textbf{ApC} & 6.78  & 6.75  & 4.97  & 2.85  
       & 4.81  & 2.51  
       & 0.60  & 0.52  & 0.39  & 0.20  
       & 0.23  & 0.15  \\
\midrule
\multirow{5}{*}{\coesmall{BERT}}
 & \textbf{AUC} & 60.62 & 61.09 & 60.94 & 60.81 
       & 62.65 & 62.40 
       & 63.02 & 63.62 & 63.40 & 62.94 
       & 64.29 & 63.75 \\
 & \textbf{MRR} & 29.31 & 29.26 & 29.31 & 29.04 
       & 30.92 & 30.75 
       & 31.23 & 31.59 & 31.38 & 30.56 
       & 32.60 & 31.58 \\
 & \textbf{N@5} & 29.71 & 29.60 & 29.65 & 29.38 
       & 31.37 & 32.44 
       & 31.79 & 32.30 & 32.16 & 31.83 
       & 33.63 & 32.43 \\
 & \textbf{\ce $\downarrow$} & \underline{22}    & \underline{23}    & \underline{33}    & \underline{38}    
       & \underline{67}    & \underline{58}    
       & \underline{353}   & \underline{404}   & \underline{505}   & \underline{640}   
       & \underline{1010}  & \underline{1010}   \\
 & \textbf{ApC} & \underline{48.27} & \underline{48.22} & \underline{33.15} & \underline{28.45} 
       & \underline{18.88} & \underline{21.37} 
       & \underline{3.69}  & \underline{3.37}  & \underline{2.65}  & \underline{2.02}  
       & \underline{1.41}  & \underline{1.36}   \\
\midrule
\multirow{5}{*}{\coesmall{MFT}}
 & \textbf{AUC} & \textbf{62.95} & \underline{62.16} & \textbf{62.95} & \underline{62.43} 
       & \textbf{64.57} & \underline{63.12} 
       & \underline{64.78} & \underline{64.88} & \underline{64.34} & \underline{65.33} 
       & \textbf{65.44} & \underline{64.53} \\
 & \textbf{MRR} & \underline{31.26} & \underline{31.00} & \textbf{31.18} & \underline{30.42} 
       & \textbf{32.60} & \underline{31.28} 
       & \underline{32.64} & \underline{32.94} & \underline{32.93} & \underline{33.29} 
       & \textbf{33.04} & \textbf{32.72} \\
 & \textbf{N@5} & \underline{32.01} & \underline{31.79} & \textbf{32.10} & \underline{30.94} 
       & \underline{33.48} & \underline{32.01} 
       & \underline{33.66} & \underline{34.00} & \underline{33.95} & \underline{34.35} 
       & \textbf{34.03} & \underline{33.58} \\
 & \textbf{\ce $\downarrow$} & \underline{22}    & \underline{23}    & \underline{33}    & \underline{38}    
       & \underline{67}    & \underline{58}    
       & \underline{353}   & \underline{404}   & \underline{505}   & \underline{640}   
       & \underline{1010}  & \underline{1010}   \\
& \textbf{ApC} & \textbf{58.86} & \textbf{52.87} & \textbf{39.24} & \textbf{32.71} 
       & \textbf{21.75} & \textbf{22.62} 
       & \textbf{4.19}  & \textbf{3.68}  & \textbf{2.84}  & \textbf{2.40}  
       & \textbf{1.53}  & \textbf{1.44} \\
\bottomrule
\end{tabular}
}

\caption{Comparison of different variants of different recommendation baselines.
}
\label{tab:big-table}

\end{table*}

\subsection{Methods and Variants}\label{subsec:baseline} 
We benchmark 5 variants of 6 widely-used news recommendation models (30 in total), including 4 matching models (i.e., NAML~\cite{wu2019neural}, LSTUR~\cite{an2019neural}, NRMS~\cite{wu2019nrms}, and BST~\cite{bst}), and 2 ranking models (i.e., DCN~\cite{wang2017deep} and DIN~\cite{zhou2018deep}. 
We categorize some of these models with~\autoref{tab:benchmark} in terms of the used user encoder and news encoder. More details please refer to~\autoref{sec:models}.

\subsection{Evaluation across Different Frameworks}

The comparative results of various models are summarized in \autoref{tab:big-table}, from which we derive the following observations.

\textbf{First}, among the five variants, the ID-based variants exhibit the poorest performance, indicating the significance of news content comprehension in news recommender systems. 
\textbf{Second}, the pretrain-finetune framework (i.e., PLMNR variants) consistently outperforms the end-to-end framework (i.e., ID-based and Text-based variants), as the pretrained knowledge enhances the understanding of news content. 
\textbf{Third}, our \model{}-based MFT variants achieve comparable performance to PLMNR variants while significantly reducing carbon emissions.
\textbf{Fourth}, the carbon emissions of BERT variants are equivalent to the corresponding MFT variants since they both serve as the initialization of the news embedding table and share the same model backbone. However, BERT variants do not learn from multi-field content nor train with recommendation-oriented tasks, resulting in news representations that primarily contain general knowledge, leading to inferior performance compared to ours.
\textbf{Fifth}, the carbon unit conversion rates of our MFT variants demonstrate state-of-the-art performance on both datasets, highlighting their success in achieving model efficiency and sustainability.

\subsection{Ablation Study}
The performance of the two self-supervised tasks is presented in \autoref{tab:ablation-task}. Based on the results, we can conclude that both the masked token prediction and field alignment tasks contribute to overall performance improvement, confirming their effectiveness.


Next, we investigate the impact of the embedding dimension on the ApC metric, which represents the unit conversion rate of carbon emissions on the AUC metric. As depicted in \autoref{fig:per}, the following observations can be made: \textbf{First}, PLMNR variants exhibit the lowest conversion efficiency. Increasing the embedding dimension leads to a quadratic growth in the number of Transformer parameters, resulting in decreased conversion efficiency. \textbf{Second}, our \model{}-based MFT variants consistently outperform others by achieving conversion rates 1.3 to 30 times higher, depending on the embedding dimensions, demonstrating their exceptional efficiency.

\subsection{Visualization}

We provide a visual representation of the performance and energy consumption of two popular news recommendation models, NRMS~\cite{wu2019nrms} and DCN~\cite{wang2017deep}, under different training schemes: ID-based, Text-based, PLM-based (PLMNR), and MFT-based, as depicted in \autoref{fig:example}. It can be observed that PLM-based variants achieve significantly better recommendation performance than text-based and ID-based variants but at the expense of much higher carbon emissions. Notably, our MFT-based variants achieve comparable recommendation performance to PLM-based variants while emitting significantly less carbon and achieving the best unit conversion rate.

\begin{table}[]
\centering
\renewcommand\arraystretch{1.2}
\setlength\tabcolsep{3pt}

\resizebox{.9\linewidth}{!}{
\begin{tabular}{lcccccc}
\toprule
& \multicolumn{3}{c}{\textbf{NRMS}} & \multicolumn{3}{c}{\textbf{DCN}} \\
\cmidrule(lr){2-4} \cmidrule(lr){5-7}
\textbf{Task} & AUC & MRR & N@5 & AUC & MRR & N@5 \\ 
\midrule
\textbf{MTP}         & 61.78 & 30.53 & 31.20 & 62.75 & 30.81 & 31.26 \\
\textbf{MTP+FA}      & \textbf{62.95} & \textbf{31.18} & \textbf{32.10} & \textbf{64.57} & \textbf{32.60} & \textbf{33.48} \\
\bottomrule
\end{tabular}
}
\caption{\label{tab:ablation-task} 
Comparison results of different combinations of MFT self-supervised tasks on the MIND-small dataset. It should be noted that the CO$_2$E metrics remain the same across all combinations since the MFT model extracts news representations with the same dimension, regardless of the number of pretraining tasks.
}
\end{table}


\begin{figure}
    \centering
    \setlength\tabcolsep{0pt}
    \resizebox{.9\linewidth}{!}{
    \begin{tabular}{m{0.25\textwidth}m{0.25\textwidth}}
    \multicolumn{2}{c}{
        \resizebox{1.0\linewidth}{!}{
            \begin{tikzpicture}
    \begin{customlegend}[
        legend columns=3,
        legend style={
            align=left,
            draw=none,
            column sep=2ex
        },
        legend entries={
            \textsc{\small{NAML}},
            \textsc{\small{LSTUR}},
            \textsc{\small{DIN}},
        }]
        \addlegendimage{olive,mark=x,solid,line legend}
        \addlegendimage{orange,mark=x,solid,line legend}
        \addlegendimage{purple,mark=x,solid,line legend}
        \end{customlegend}
\end{tikzpicture}
        }
    } \\
    \begin{subfigure}{0.25\textwidth}
        \resizebox{1.0\linewidth}{!}{
            \begin{tikzpicture}
    \begin{axis}[
        xtick={1,2,3,4},
        xticklabels={ID-based, Text-Based, PLMNR, MFT-news},
        line width=0.35mm,
        legend style={at={(axis cs:5,0)},anchor=south west},
        ymin=0,
        ymax=60,
        ylabel=ApC,
    ]

        \addplot[
            purple,
            mark=x,
        ]
            coordinates{
                (1, 0.68) (2, 24.14) (3, 6.78) (4, 58.86)
            }; 
            
        \addplot[
            orange,
            mark=x,
        ]
            coordinates{
                (1, 5.20) (2, 19.43) (3, 6.75) (4, 52.87)
            }; 
            
        \addplot[
            olive,
            mark=x,
        ]
            coordinates{
                (1, 10.26) (2, 16.33) (3, 2.51) (4, 22.62)
            }; 

    \end{axis}

\end{tikzpicture}
        }
        \caption{\label{fig:dim-64}64 dimension}
    \end{subfigure} & 
    \begin{subfigure}{0.25\textwidth}
        \resizebox{1.0\linewidth}{!}{
            \begin{tikzpicture}
    \begin{axis}[
        xtick={1,2,3,4},
        xticklabels={ID-based, Text-Based, PLMNR, MFT-news},
        line width=0.35mm,
        legend style={at={(axis cs:5,0)},anchor=south west},
        ymin=0,
        ymax=60,
        ylabel=ApC,
    ]

        \addplot[
            purple,
            mark=x,
        ]
            coordinates{
                (1, 3.12) (2, 14.51) (3, 1.59) (4, 46.64)
            }; 
            
        \addplot[
            orange,
            mark=x,
        ]
            coordinates{
                (1, 2.85) (2, 12.28) (3, 1.42) (4, 52.70)
            }; 
            
        \addplot[
            olive,
            mark=x,
        ]
            coordinates{
                (1, 8.34) (2, 11.90) (3, 0.54) (4, 16.32)
            }; 

    \end{axis}

\end{tikzpicture}
        }
        \caption{\label{fig:dim-768}768 dimension}
    \end{subfigure}
    \end{tabular}
    }
    
    \caption{\label{fig:per} Unit conversion rate of carbon emission on NAML, LSTUR, and DIN models with variants on the MIND-small dataset.}
\end{figure}
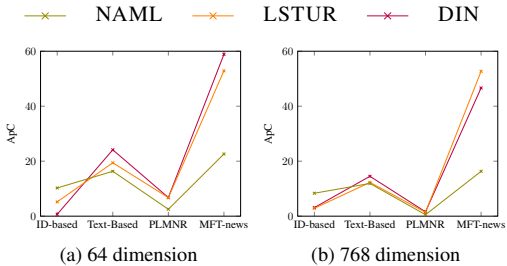
\section{Related Work}

\textbf{Green AI}~\cite{schwartz2020green} and Green Learning~\cite{kuo2022green} are a relatively new area of research, with a growing number of studies~\cite{wilson2020urban,przybyla2022using,thirukovalluru2021scaling} exploring ways to reduce the environmental impact of artificial intelligence. One line of research has focused on developing more energy-efficient hardware~\cite{mudigere2022software} for deep learning, such as specialized chips and processors optimized for deep learning workloads. Another line of research has focused on developing more energy-efficient algorithms for deep learning, such as decoupling~\cite{vanpre,shi2022layerconnect}, distillation~\cite{sanh2019distilbert}, and quantization~\cite{diao2021generalized}, which are all proposed to decrease the energy consumption and carbon emissions.

\definecolor{zhucao}{rgb}{0.65, 0.25, 0.21}
\definecolor{fuguang}{rgb}{0.94, 0.76, 0.64}
\definecolor{tushoulan}{rgb}{0.50, 0.76, 0.89}
\definecolor{qingdai}{rgb}{0.51, 0.72, 0.61}

\begin{figure}
\centering
\setlength\tabcolsep{0pt}

\resizebox{0.75\linewidth}{!}{
\begin{tabular}{c}

\begin{tikzpicture}

    \begin{groupplot}[
        group style={
            group name=my fancy plots,
            group size=1 by 2,
            xticklabels at=edge bottom,
            vertical sep=0pt
        },
        width=8.5cm,
        xtick={0, 2000, 4000, 6000},
        xticklabels={0, 2, 4, 6},
        xmin=0,
        xmax=7000,
        xlabel=Carbon Emission (Kg),
        ylabel=AUC,
        xmajorgrids=true,
        ymajorgrids=true,
        grid style=dashed,
        legend columns=2,
        legend style={
            font=\normalsize,
            legend cell align=left,
        },
        legend pos=south east,
    ]
    
    \nextgroupplot[
        ymin=63,
        ymax=66,
        ytick={64, 66},
        axis x line=top, 
        axis y discontinuity=parallel,
        height=6.0cm,
        xlabel=,
        y label style={at={(-0.1,0.1)}},
    ]
        \addplot[color=fuguang, only marks, style={mark=*, fill=fuguang,mark size=4pt}] coordinates {(1010,64.12)};\label{plot:nrms}
        \addplot[color=fuguang, only marks, style={mark=triangle*, fill=fuguang,mark size=5pt}] coordinates {(1684,63.88)};\label{plot:dcn}
        
        \addplot[color=qingdai, only marks, style={mark=*, 
        fill=qingdai,mark size=4pt}] coordinates {(4043,65.57)};\label{plot:nrms-plm}
        \addplot[color=qingdai, only marks, style={mark=triangle*, fill=qingdai,mark size=5pt}] coordinates {(6570,65.42)};\label{plot:dcn-plm}
        
        \addplot[color=zhucao, only marks, style={mark=*, fill=zhucao,mark size=4pt}] coordinates {(505,64.34)};\label{plot:nrms-gnrs}
        \addplot[color=zhucao, only marks, style={mark=triangle*, fill=zhucao,mark size=5pt}] coordinates {(1010,65.44)};\label{plot:dcn-gnrs}
    
    \nextgroupplot[
        ymin=56,ymax=58,
        ytick={56, 58},
        axis x line=bottom,
        height=4.0cm,
        ylabel=,
    ]
        \addplot[color=tushoulan, only marks, style={mark=*, fill=tushoulan,mark size=4pt}] coordinates {(471,57.59)}; \label{plot:nrms-id}
        \addplot[color=tushoulan, only marks, style={mark=triangle*, fill=tushoulan,mark size=5pt}] coordinates {(791,57.41)};\label{plot:dcn-id}
    \end{groupplot}
    
        \node[draw,fill=white,inner sep=4pt,above left=0.5em] at (6.7, -2.4) {
        \setlength\tabcolsep{2pt}
        \begin{tabular}{ccl}
        $\texttt{NRMS}$ & $\texttt{DCN}$ \\
        \ref{plot:nrms-id} & \ref{plot:dcn-id} & $\texttt{ID-based}$\\
        \ref{plot:nrms} & \ref{plot:dcn} & $\texttt{Text-based}$\\
        \ref{plot:nrms-plm} & \ref{plot:dcn-plm} & $\texttt{PLM-based}$ \\
        \ref{plot:nrms-gnrs} & \ref{plot:dcn-gnrs} & $\texttt{\model{} \small{(ours)}}$
        \end{tabular}
        };
\end{tikzpicture}
\end{tabular}
}
\caption{\label{fig:example} AUC and carbon emission (\ce) of two popular news recommendation methods NRMS and DCN under different training frameworks on the MIND-large dataset. Our proposed \model{} achieves the best tradeoff between performance and energy consumption. 
}
\end{figure}
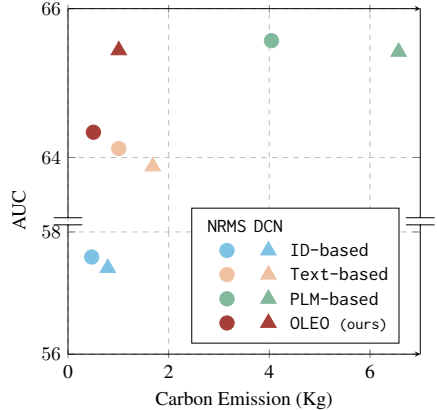

\textbf{News Recommendation.} 
Recent news recommendation methods~\cite{wu2019neural,an2019neural,wu2019nrms} apply traditional natural language processing model, such as convolutional neural network (CNN)~\cite{kim2014convolutional}, GRU~\cite{chung2014empirical}, or Attention mechanism~\cite{vaswani2017attention} to capture news content knowledge or user behavior history (i.e., clicked news). These end-to-end models are trained with recommendation task and typically have limited ability in content understanding. Most recently, PLMs such as BERT~\cite{devlin2018bert} have shown their success in news recommendation~\cite{wu2021empowering,zhang2021unbert,li2022miner,bi2022mtrec}. However, these PLM-based news encoders are trained with hundred millions of click data, and such coupled framework will result in huge computational cost and consequent giant carbon footprint.



In this paper, we aim to decouple the news recommender system into news representation learning and downstream recommendation task learning. The learned news representation vectors can be used for multiple recommender systems including ranking and matching. The decoupled design greatly reduces the computational cost while achieving competitive performance compared with the state-of-the-art baselines.
\section{Conclusion}

In this work, we propose a novel news recommendation framework \model{}, consisting of two distinct components: news representation learning and downstream recommendation tasks. Compared with the state-of-the-art PLM-based framework, our framework demonstrates significantly improved energy efficiency and attains the highest unit conversion rate of carbon emission. Within \model{} framework, we develop a multi-field Transformer instantiation that effectively learns news representations by leveraging multi-field news content. Extensive experiments demonstrate that our \model{}-based MFT model strikes a superior balance between quality and efficiency performance. We hope that our work will inspire more explorations into the efficient recommender system domain.

\section*{Limitations}

In terms of limitations, there are two aspects that need to be addressed. Firstly, our evaluation has been conducted primarily on the MIND dataset, and it would be beneficial to include evaluations on additional real-world datasets, such as the recently proposed PENS dataset~\cite{ao2021pens}. 
Furthermore, while we have compared our approach against several baseline models, there are still other well-established news recommendation baselines, such as DKN~\cite{wang2018dkn} and FIM~\cite{wang2020fine}, that could be included for a more thorough comparison. 



\bibliography{GNR}

\begin{thebibliography}{37}
\expandafter\ifx\csname natexlab\endcsname\relax\def\natexlab#1{#1}\fi

\bibitem[{An et~al.(2019)An, Wu, Wu, Zhang, Liu, and Xie}]{an2019neural}
Mingxiao An, Fangzhao Wu, Chuhan Wu, Kun Zhang, Zheng Liu, and Xing Xie. 2019.
\newblock Neural news recommendation with long-and short-term user
  representations.
\newblock In \emph{Proceedings of the 57th Annual Meeting of the Association
  for Computational Linguistics}, pages 336--345.

\bibitem[{Ao et~al.(2021)Ao, Wang, Luo, Qiao, He, and Xie}]{ao2021pens}
Xiang Ao, Xiting Wang, Ling Luo, Ying Qiao, Qing He, and Xing Xie. 2021.
\newblock Pens: A dataset and generic framework for personalized news headline
  generation.
\newblock In \emph{Proceedings of the 59th Annual Meeting of the Association
  for Computational Linguistics and the 11th International Joint Conference on
  Natural Language Processing (Volume 1: Long Papers)}, pages 82--92.

\bibitem[{Bahdanau et~al.(2014)Bahdanau, Cho, and Bengio}]{bahdanau2014neural}
Dzmitry Bahdanau, Kyunghyun Cho, and Yoshua Bengio. 2014.
\newblock Neural machine translation by jointly learning to align and
  translate.
\newblock \emph{arXiv preprint arXiv:1409.0473}.

\bibitem[{Bi et~al.(2022)Bi, Li, Shang, Jiang, Liu, and Yang}]{bi2022mtrec}
Qiwei Bi, Jian Li, Lifeng Shang, Xin Jiang, Qun Liu, and Hanfang Yang. 2022.
\newblock Mtrec: Multi-task learning over bert for news recommendation.
\newblock In \emph{Findings of the Association for Computational Linguistics:
  ACL 2022}, pages 2663--2669.

\bibitem[{Chen et~al.(2019)Chen, Zhao, Li, Huang, and Ou}]{bst}
Qiwei Chen, Huan Zhao, Wei Li, Pipei Huang, and Wenwu Ou. 2019.
\newblock Behavior sequence transformer for e-commerce recommendation in
  alibaba.
\newblock In \emph{Proceedings of the 1st International Workshop on Deep
  Learning Practice for High-Dimensional Sparse Data}, pages 1--4.

\bibitem[{Chung et~al.(2014)Chung, Gulcehre, Cho, and
  Bengio}]{chung2014empirical}
Junyoung Chung, Caglar Gulcehre, KyungHyun Cho, and Yoshua Bengio. 2014.
\newblock Empirical evaluation of gated recurrent neural networks on sequence
  modeling.
\newblock \emph{arXiv preprint arXiv:1412.3555}.

\bibitem[{Devlin et~al.(2018)Devlin, Chang, Lee, and
  Toutanova}]{devlin2018bert}
Jacob Devlin, Ming-Wei Chang, Kenton Lee, and Kristina Toutanova. 2018.
\newblock Bert: Pre-training of deep bidirectional transformers for language
  understanding.
\newblock \emph{arXiv preprint arXiv:1810.04805}.

\bibitem[{Diao et~al.(2021)Diao, Kleyko, Rabaey, and
  Olshausen}]{diao2021generalized}
Cameron Diao, Denis Kleyko, Jan~M Rabaey, and Bruno~A Olshausen. 2021.
\newblock Generalized learning vector quantization for classification in
  randomized neural networks and hyperdimensional computing.
\newblock In \emph{2021 International Joint Conference on Neural Networks
  (IJCNN)}, pages 1--9. IEEE.

\bibitem[{Jing and Tian(2019)}]{VisualSelfSupervision}
Longlong Jing and Yingli Tian. 2019.
\newblock Self-supervised visual feature learning with deep neural networks:
  {A} survey.
\newblock \emph{CoRR}, abs/1902.06162.

\bibitem[{Kim(2014)}]{kim2014convolutional}
Yoon Kim. 2014.
\newblock Convolutional neural networks for sentence classification.
\newblock \emph{arXiv preprint arXiv:1408.5882}.

\bibitem[{Krizhevsky et~al.(2012)Krizhevsky, Sutskever, and
  Hinton}]{krizhevsky2012imagenet}
Alex Krizhevsky, Ilya Sutskever, and Geoffrey~E Hinton. 2012.
\newblock Imagenet classification with deep convolutional neural networks.
\newblock \emph{Advances in neural information processing systems},
  25:1097--1105.

\bibitem[{Kuo and Madni(2022)}]{kuo2022green}
C-C~Jay Kuo and Azad~M Madni. 2022.
\newblock Green learning: Introduction, examples and outlook.
\newblock \emph{Journal of Visual Communication and Image Representation}, page
  103685.

\bibitem[{Lacoste et~al.(2019)Lacoste, Luccioni, Schmidt, and
  Dandres}]{lacoste2019quantifying}
Alexandre Lacoste, Alexandra Luccioni, Victor Schmidt, and Thomas Dandres.
  2019.
\newblock Quantifying the carbon emissions of machine learning.
\newblock \emph{arXiv preprint arXiv:1910.09700}.

\bibitem[{Li et~al.(2022)Li, Zhu, Bi, Cai, Shang, Dong, Jiang, and
  Liu}]{li2022miner}
Jian Li, Jieming Zhu, Qiwei Bi, Guohao Cai, Lifeng Shang, Zhenhua Dong, Xin
  Jiang, and Qun Liu. 2022.
\newblock Miner: Multi-interest matching network for news recommendation.
\newblock In \emph{Findings of the Association for Computational Linguistics:
  ACL 2022}, pages 343--352.

\bibitem[{Liu et~al.(2022)Liu, Zhu, Dai, and Wu}]{prec}
Qijiong Liu, Jieming Zhu, Quanyu Dai, and Xiaoming Wu. 2022.
\newblock \href {https://aclanthology.org/2022.coling-1.249} {Boosting deep
  {CTR} prediction with a plug-and-play pre-trainer for news recommendation}.
\newblock In \emph{Proceedings of the 29th International Conference on
  Computational Linguistics}, pages 2823--2833, Gyeongju, Republic of Korea.
  International Committee on Computational Linguistics.

\bibitem[{Mudigere et~al.(2022)Mudigere, Hao, Huang, Jia, Tulloch, Sridharan,
  Liu, Ozdal, Nie, Park et~al.}]{mudigere2022software}
Dheevatsa Mudigere, Yuchen Hao, Jianyu Huang, Zhihao Jia, Andrew Tulloch,
  Srinivas Sridharan, Xing Liu, Mustafa Ozdal, Jade Nie, Jongsoo Park, et~al.
  2022.
\newblock Software-hardware co-design for fast and scalable training of deep
  learning recommendation models.
\newblock In \emph{Proceedings of the 49th Annual International Symposium on
  Computer Architecture}, pages 993--1011.

\bibitem[{Przyby{\l}a and Shardlow(2022)}]{przybyla2022using}
Piotr Przyby{\l}a and Matthew Shardlow. 2022.
\newblock Using nlp to quantify the environmental cost and diversity benefits
  of in-person nlp conferences.
\newblock In \emph{Findings of the Association for Computational Linguistics:
  ACL 2022}, pages 3853--3863.

\bibitem[{Radford et~al.(2021)Radford, Kim, Hallacy, Ramesh, Goh, Agarwal,
  Sastry, Askell, Mishkin, Clark, Krueger, and Sutskever}]{CLIP}
Alec Radford, Jong~Wook Kim, Chris Hallacy, Aditya Ramesh, Gabriel Goh,
  Sandhini Agarwal, Girish Sastry, Amanda Askell, Pamela Mishkin, Jack Clark,
  Gretchen Krueger, and Ilya Sutskever. 2021.
\newblock Learning transferable visual models from natural language
  supervision.
\newblock In \emph{Proceedings of the 38th International Conference on Machine
  Learning ({ICML})}, pages 8748--8763.

\bibitem[{Sanh et~al.(2019)Sanh, Debut, Chaumond, and
  Wolf}]{sanh2019distilbert}
Victor Sanh, Lysandre Debut, Julien Chaumond, and Thomas Wolf. 2019.
\newblock Distilbert, a distilled version of bert: smaller, faster, cheaper and
  lighter.
\newblock \emph{arXiv preprint arXiv:1910.01108}.

\bibitem[{Schwartz et~al.(2020)Schwartz, Dodge, Smith, and
  Etzioni}]{schwartz2020green}
Roy Schwartz, Jesse Dodge, Noah~A Smith, and Oren Etzioni. 2020.
\newblock Green ai.
\newblock \emph{Communications of the ACM}, 63(12):54--63.

\bibitem[{Shi et~al.(2022)Shi, Zhang, Wang, Wang, Zheng, and
  Sakai}]{shi2022layerconnect}
Haoxiang Shi, Rongsheng Zhang, Jiaan Wang, Cen Wang, Yinhe Zheng, and Tetsuya
  Sakai. 2022.
\newblock Layerconnect: Hypernetwork-assisted inter-layer connector to enhance
  parameter efficiency.
\newblock In \emph{Proceedings of the 29th International Conference on
  Computational Linguistics}, pages 3120--3126.

\bibitem[{Thirukovalluru et~al.(2021)Thirukovalluru, Monath, Shridhar, Zaheer,
  Sachan, and McCallum}]{thirukovalluru2021scaling}
Raghuveer Thirukovalluru, Nicholas Monath, Kumar Shridhar, Manzil Zaheer,
  Mrinmaya Sachan, and Andrew McCallum. 2021.
\newblock Scaling within document coreference to long texts.
\newblock In \emph{Findings of the Association for Computational Linguistics:
  ACL-IJCNLP 2021}, pages 3921--3931.

\bibitem[{van Cauter()}]{vanpre}
Zeno van Cauter.
\newblock Pre-training large nlp-models by utilizing low-resource
  nlp-pipelines.

\bibitem[{Vaswani et~al.(2017)Vaswani, Shazeer, Parmar, Uszkoreit, Jones,
  Gomez, Kaiser, and Polosukhin}]{vaswani2017attention}
Ashish Vaswani, Noam Shazeer, Niki Parmar, Jakob Uszkoreit, Llion Jones,
  Aidan~N Gomez, Lukasz Kaiser, and Illia Polosukhin. 2017.
\newblock Attention is all you need.
\newblock \emph{arXiv preprint arXiv:1706.03762}.

\bibitem[{Wang et~al.(2020)Wang, Wu, Liu, and Xie}]{wang2020fine}
Heyuan Wang, Fangzhao Wu, Zheng Liu, and Xing Xie. 2020.
\newblock Fine-grained interest matching for neural news recommendation.
\newblock In \emph{Proceedings of the 58th Annual Meeting of the Association
  for Computational Linguistics}, pages 836--845.

\bibitem[{Wang et~al.(2018)Wang, Zhang, Xie, and Guo}]{wang2018dkn}
Hongwei Wang, Fuzheng Zhang, Xing Xie, and Minyi Guo. 2018.
\newblock Dkn: Deep knowledge-aware network for news recommendation.
\newblock In \emph{Proceedings of the 2018 world wide web conference}, pages
  1835--1844.

\bibitem[{Wang et~al.(2017)Wang, Fu, Fu, and Wang}]{wang2017deep}
Ruoxi Wang, Bin Fu, Gang Fu, and Mingliang Wang. 2017.
\newblock Deep \& cross network for ad click predictions.
\newblock In \emph{Proceedings of the ADKDD'17}, pages 1--7.

\bibitem[{Wilson et~al.(2020)Wilson, Magdy, McGillivray, Garimella, and
  Tyson}]{wilson2020urban}
Steven Wilson, Walid Magdy, Barbara McGillivray, Kiran Garimella, and Gareth
  Tyson. 2020.
\newblock Urban dictionary embeddings for slang nlp applications.
\newblock In \emph{Proceedings of the Twelfth Language Resources and Evaluation
  Conference}, pages 4764--4773.

\bibitem[{Wolf et~al.(2020)Wolf, Debut, Sanh, Chaumond, Delangue, Moi, Cistac,
  Rault, Louf, Funtowicz, Davison, Shleifer, von Platen, Ma, Jernite, Plu, Xu,
  Scao, Gugger, Drame, Lhoest, and Rush}]{wolf-etal-2020-transformers}
Thomas Wolf, Lysandre Debut, Victor Sanh, Julien Chaumond, Clement Delangue,
  Anthony Moi, Pierric Cistac, Tim Rault, Rémi Louf, Morgan Funtowicz, Joe
  Davison, Sam Shleifer, Patrick von Platen, Clara Ma, Yacine Jernite, Julien
  Plu, Canwen Xu, Teven~Le Scao, Sylvain Gugger, Mariama Drame, Quentin Lhoest,
  and Alexander~M. Rush. 2020.
\newblock \href {https://www.aclweb.org/anthology/2020.emnlp-demos.6}
  {Transformers: State-of-the-art natural language processing}.
\newblock In \emph{Proceedings of the 2020 Conference on Empirical Methods in
  Natural Language Processing: System Demonstrations}, pages 38--45, Online.
  Association for Computational Linguistics.

\bibitem[{Wu et~al.(2019{\natexlab{a}})Wu, Wu, An, Huang, Huang, and
  Xie}]{wu2019neural}
Chuhan Wu, Fangzhao Wu, Mingxiao An, Jianqiang Huang, Yongfeng Huang, and Xing
  Xie. 2019{\natexlab{a}}.
\newblock Neural news recommendation with attentive multi-view learning.
\newblock \emph{IJCAI}.

\bibitem[{Wu et~al.(2019{\natexlab{b}})Wu, Wu, An, Huang, Huang, and
  Xie}]{wu2019npa}
Chuhan Wu, Fangzhao Wu, Mingxiao An, Jianqiang Huang, Yongfeng Huang, and Xing
  Xie. 2019{\natexlab{b}}.
\newblock Npa: neural news recommendation with personalized attention.
\newblock In \emph{Proceedings of the 25th ACM SIGKDD International Conference
  on Knowledge Discovery \& Data Mining}, pages 2576--2584.

\bibitem[{Wu et~al.(2019{\natexlab{c}})Wu, Wu, Ge, Qi, Huang, and
  Xie}]{wu2019nrms}
Chuhan Wu, Fangzhao Wu, Suyu Ge, Tao Qi, Yongfeng Huang, and Xing Xie.
  2019{\natexlab{c}}.
\newblock Neural news recommendation with multi-head self-attention.
\newblock In \emph{Proceedings of the 2019 Conference on Empirical Methods in
  Natural Language Processing and the 9th International Joint Conference on
  Natural Language Processing (EMNLP-IJCNLP)}, pages 6390--6395.

\bibitem[{Wu et~al.(2021)Wu, Wu, Qi, and Huang}]{wu2021empowering}
Chuhan Wu, Fangzhao Wu, Tao Qi, and Yongfeng Huang. 2021.
\newblock Empowering news recommendation with pre-trained language models.
\newblock \emph{SIGIR}.

\bibitem[{Wu et~al.(2020)Wu, Qiao, Chen, Wu, Qi, Lian, Liu, Xie, Gao, Wu
  et~al.}]{wu2020mind}
Fangzhao Wu, Ying Qiao, Jiun-Hung Chen, Chuhan Wu, Tao Qi, Jianxun Lian,
  Danyang Liu, Xing Xie, Jianfeng Gao, Winnie Wu, et~al. 2020.
\newblock Mind: A large-scale dataset for news recommendation.
\newblock In \emph{Proceedings of the 58th Annual Meeting of the Association
  for Computational Linguistics}, pages 3597--3606.

\bibitem[{Yang et~al.(2019)Yang, Wang, Liu, Liu, Lyu, Wu, She, and
  Li}]{EnhancingKnowledge}
An~Yang, Quan Wang, Jing Liu, Kai Liu, Yajuan Lyu, Hua Wu, Qiaoqiao She, and
  Sujian Li. 2019.
\newblock Enhancing pre-trained language representations with rich knowledge
  for machine reading comprehension.
\newblock In \emph{Proceedings of the 57th Conference of the Association for
  Computational Linguistics (ACL)}, pages 2346--2357.

\bibitem[{Zhang et~al.(2021)Zhang, Li, Jia, Wang, Zhu, Wang, and
  He}]{zhang2021unbert}
Qi~Zhang, Jingjie Li, Qinglin Jia, Chuyuan Wang, Jieming Zhu, Zhaowei Wang, and
  Xiuqiang He. 2021.
\newblock Unbert: User-news matching bert for news recommendation.
\newblock In \emph{Proceedings of the Thirtieth International Joint Conference
  on Artificial Intelligence (IJCAI)}, pages 3356--3362.

\bibitem[{Zhou et~al.(2018)Zhou, Zhu, Song, Fan, Zhu, Ma, Yan, Jin, Li, and
  Gai}]{zhou2018deep}
Guorui Zhou, Xiaoqiang Zhu, Chenru Song, Ying Fan, Han Zhu, Xiao Ma, Yanghui
  Yan, Junqi Jin, Han Li, and Kun Gai. 2018.
\newblock Deep interest network for click-through rate prediction.
\newblock In \emph{Proceedings of the 24th ACM SIGKDD international conference
  on knowledge discovery \& data mining}, pages 1059--1068.

\end{thebibliography}
\bibliographystyle{acl_natbib}

\appendix



\section{Details of Methods and Variants}

\label{sec:models}

\textbf{End-to-End Framework.} \textit{1) ID-based variants.} They directly ignore news content information. BST, DCN, and DIN are original ID-based methods. ID-NAML, ID-LSTUR, and ID-NRMS methods remove the news encoders of NAML, LSTUR, and NRMS, respectively, and directly connect the trainable news embedding to their user encoders. \textit{2) Text-based variants.} NAML, LSTUR, NRMS are original Text-based methods. Text-DCN, Text-DIN, and Text-BST use a simple average pooling layer as their news encoder to fuse content information.

\textbf{Pretrain-Finetune Framework.} \textit{PLM-empowered Variants.} Following PLMNR~\cite{wu2021empowering}, which proposes to replace original news encoder with PLMs, we construct PLMNR-NAML, PLMNR-LSTUR, PLMNR-NRMS, PLMNR-BST, PLMNR-DCN, PLMNR-DIN methods, where the latter three methods directly use PLMs as their news encoder. An additive attention network is added above the PLM to fuse sequential information into a unified news representation.

\textbf{Proposed \model{} Framework.} \textit{1) BERT empowered id-based variants.} After BERT's encoding news content, we perform an average pooling on BERT's output and obtain news representations. The representations are applied to initialize the above ID-based variants. \textit{2) MFT-news empowered id-based variants.} We pretrain the MFT model with news data, and obtain news representations after pretrained MFT's encoding. The representations are applied to initialize the above ID-based variants.


\end{document}